\documentclass[preprint,onecolumn,...]{revtex4-1}

\usepackage{amsmath}
\usepackage{amsfonts}
\usepackage{graphicx}
\usepackage{hyperref}
\usepackage{mathbbol}
\usepackage{setspace}

\def\be{\begin{equation}}
\def\ee{\end{equation}}
\def\bea{\begin{eqnarray}}
\def\eea{\end{eqnarray}}

\def\f{\frac}

\def\o{\omega}
\def\l{\left}
\def\r{\right}

\def\no{\nonumber}
\def\d{{\rm d}}

\begin{document}


\title{Born approximation in linear-time invariant system}
\author{Burin Gumjudpai}
\email{buring@nu.ac.th} \affiliation{The Institute for Fundamental Study ``The Tah Poe Academia Institute" \\ Naresuan University, Phitsanulok 65000, Thailand}

\begin{abstract}
Linear-time invariant (LTI) oscillation systems such as forced mechanical vibration,
series RLC and parallel RLC circuits can be solved by using simplest initial conditions or
employing of Green's function of which knowledge of initial condition
of the force term is needed. Here we show a mathematical connection of the LTI system and the Helmholtz equation
form of the time-independent Schr\"{o}dinger equation in quantum mechanical scattering problem. We apply Born approximation in quantum mechanics
 to obtain LTI general solution in form of infinite Born series which can be expressed as a series of one-dimensional Feynman graphs.
 Conditions corresponding to the approximation are given for the case of harmonic driving force. The Born series of the harmonic forced oscillation case are derived by directly applying the approximation to the LTI system or by transforming the LTI system to Helmholtz equation prior to doing the approximation.
\end{abstract}

\date{\today}

\maketitle

\section{Introduction}

\label{sec:introduction}
Physical oscillation is an important phenomena in physics and engineering. It is typically modeled as a linear ordinary differential equation considering the non-linear effect is negligibly small. For a system with time-independency of its physical properties, the system is known as linear-time invariant (LTI) system. Typical examples of the LTI system are dynamical oscillation in spring and in RLC circuits, both series and parallel. Important properties of the system is the principle of superposition. When the LTI system is given an input signal, $F(t)$ as viewed in engineering, or equivalently the system is given an external driving force as viewed in physics, the differential equation of the system becomes inhomogeneous. General solution of the linear inhomogeneous ordinary differential equation is a sum of its complementary and particular solutions.  When applying a periodic input signal, particular solution can be obtained with the help of Fourier series. For arbitrary form of the input signal, the particular solution can be found with Green's function method (see, e.g. Barton \cite{bar}) of which knowledge of initial condition of the input signal $F(t_0)$ is necessary for solving the system.
Green's function of the second-order LTI system is well-known and the initial input signal (initial value of the driving force) in the RLC system can be setup experimentally. This enables us to find particular solution for an arbitrary shape of the input signal.

In non-relativistic quantum mechanics, particle with mass $m$ scattered from a potential source can be described by time-independent Schr\"{o}dinger equation written in form of Helmholtz equation. Distinct aspect between the LTI and the Helmholtz equation is that there is a first-order derivative term (damping term) in the LTI but not in the Helmholtz equation. Common aspects are such as they are second-order linear equation with source term. The source term, $Q({\bf r})$ in the Helmholtz equation is constructed from multiplication of the potential, $V({\bf r})$ and the wave function $\psi({\bf r})$ which is a general solution of the system, making $Q({\bf r})$ non-designable. This is unlike the source term $F(t)$ of the LTI system, which could be designed or determined by experimental setup.  In solving the Helmholtz equation, the general solution, $\psi({\bf r})$  has its own boundary value embedded in itself, hence can not be solved normally. Typical method of solving the problem is to employ Born approximation, that is to say, finding the particular solution perturbatively. Under the approximation, we do not need to know exact form of the general solution ($\psi({\bf r})$) before hand, but instead we can approximately use the complementary solution as a general solution at the boundary, assuming the complementary solution is not much altered by the source term at the boundary. This makes the solution recursive and hence perturbative.

In this work, we present a direct usage of Born approximation in the LTI system given some approximation conditions for the Born approximation to be valid.
In addition, alternative to directly applying Born approximation to the LTI system, we transform the LTI system into Helmholtz equation form and then apply the Born approximation to it. Both ways give us general solution as infinite series which can be expressed as one-dimensional Feynman graph.  In Sec. \ref{sec1} we briefly introduce linear time-invariant system and its Green's function before, in Sec. \ref{sec2},  introducing the Helmholtz equation and its solution obtained from Green's function method. Born approximation is applied to the system quoting standard textbooks (e.g. Griffiths \cite{gf} or
Schiff \cite{S}). The second-order LTI system and the time-independent Schr\"{o}dinger equation written in form of Helmholtz equation are compared to each other. In Sec. \ref{sec3} the Born approximation is applied directly to the LTI system and the approximation conditions for the Born approximation in LTI system to be valid are found. In Sec. \ref{sec4}, last part of this work is to transform the LTI system to the Helmholtz equation form and then applying the Born approximation to the system.

\section{Linear-Time Invariant system}  \label{sec1}
In dynamics, equation of motion can be viewed as a linear operator $ \hat{\mathcal{L}}$ acting on a function $y(t)$ as
\be
\hat{\mathcal{L}}   y(t)  \:=\:  {\mathcal{F}}(t)   \,, \label{e1}
\ee
where $t$ is an independent variable.   Linear property of the operator implies principle of superposition, i.e.
\be
\hat{\mathcal{L}}   \l[ \sum_{i=1}^{N} c_i y_i(t) \r] \: =\:   \sum_{i=1}^{N}  c_i {\mathcal{F}}_i (t) \,,
\ee
for $N$ number of solutions.
As in (\ref{e1}), any solution $y(t)$ and any inhomogeneous part $F(t)$  can be expressed as
\be
y(t) \: =\:    \sum_{i=1}^{N} c_i y_i(t)\,, \;\;\;\;\;\;\;\;\;  {\mathcal{F}}(t) \:=\: \sum_{i=1}^{N}  c_i {\mathcal{F}}_i (t) \,.
\ee
A system considered here is the linear time-invariant system (LTI) with linear operator in form of
\be
\hat{\mathcal{L}}  \;=\;  a_n \f{\d^n}{\d t^n} \:  +\: a_{n-1} \f{\d^{n-1}}{\d t^{n-1}} \:+\: \ldots \:+\:  a_1 \f{\d}{\d t} \:+\: a_0 \,,
\ee
where $a_1, \ldots a_n$ are time-independent physical parameters. For $n=2$, the second-order LTI system is written as
\be
 \l[ \f{\d^2}{\d t^2}   \:  + \: \f{a_{1}}{a_2} \f{\d}{\d t}   \: + \: \f{a_0}{a_2}\r]  y(t) \;=\; \f{{\mathcal{F}}(t)}{a_2}   \;\equiv\; F(t)\,.  \label{e6}
\ee
Defined here $  {a_1}/{a_2} \equiv 2 \beta$ and $ {a_0}/{a_2} \equiv \omega_0^2 $. Three typical examples of the system are
\begin{itemize}
\item{{\it Mechanical linear oscillation}:} the function $y(t)$ is the displacement $x(t)$, $a_2 = m$, $2\beta = b/m$, $\;\omega_0^2 = s/m$
where $m$ is mass of the oscillator, $b$ is resistance in unit of kg$\cdot$sec$^{-1}$ and $s$ is Hooke's spring constant. ${\mathcal{F}}$ is external driving force hence $F \equiv {\mathcal{F}}/a_2$ is force per unit mass.
\item{{\it Series RLC circuit}:}
in physics, $y(t)$ is electrical charge $q(t)$, $a_2 = L$, $\;2\beta = R/L$, $\omega_0^2 = 1/(LC)$ and ${\mathcal{F}}(t)
$ is the voltage. In electrical engineering, it is typical to consider $y(t)$ as electrical current $i(t)$,
${\mathcal{F}}(t)$ is time derivative of voltage and the rest are the same.
\item{{\it Parallel  RLC circuit}:}
$y(t)$ is voltage $v(t)$,  $a_2 = C$,    $\;2\beta = 1/(RC)$, $\omega_0^2 = 1/(LC)$, ${\mathcal{F}}(t)$ is time derivative of the current.
\end{itemize}
The damping  $\beta$ is Neper frequency, $L$ is inductance, $C$ is capacitance and $R$ is electrical resistance.
Conclusion of analogy between these three systems is presented in Table {\ref{tab1}}. These are well known and we refer to physics or engineering textbooks such as  Kelly \cite{Kelly}, Cha and Molinder \cite{Cha} and Main \cite{Main}.
The equation (\ref{e6}) is hence
 \be
 \f{\d^2 y}{\d t^2}   \:  + \: 2 \beta \f{\d y}{\d t}   \: + \: \omega_0^2 y \;=\; F(t)  \,,  \label{e7}
\ee
 where the inertia (e.g. mass, $m$) is absorbed into the function $F(t)$. This differential equation has  general solution,
 \be
y(t)  \;=\; y_{\rm c}(t)  \:+\:  y_{\rm p}(t) \,.   \label{e8}
\ee
 The complementary solution $y_{\rm c}$ is a solution of homogeneous system (when $F(t) = 0$) and the particular solution is of the inhomogeneous case, i.e. non-zero $F(t)$.
As it is well-known that for harmonic $F(t)$, e.g.
 $
 F(t) \;=\; F_0 \cos (\o t - \phi)
 $
 the particular solution takes the form
 \be
 y_{\rm p}(t) \;=\;   \f{F_0}{\sqrt{ (\o_0^2 -\o^2)^2   +   4 \o^2 \beta^2}}   \cos (\o t - \phi - \xi)  \,,
 \ee
with $\xi = \arctan \l[ 2\o\beta /(\o_0^2  - \o^2)\r]$ and $\phi$ is the initial phase.
Here the $y_{\rm p}(t)$ represents steady-state solution. Indeed for function $F(t)$ with any periodic forms, Fourier series method can help finding $y_{\rm p}(t)$.  In case of arbitrary form of $F(t)$, technique of Green's function method is employed. This is because when $F(t)$ is in form of Dirac's delta function $\delta(t)$, i.e.
 \be
\l[  \f{\d^2 }{\d t^2}   \:  + \: 2 \beta \f{\d }{\d t}   \: + \: \omega_0^2 \r] G(t)\;=\; \delta(t)   \,,   \label{e11}
 \ee
general solution is the Green's function, $G(t)$. Hence one can consider arbitrary force function as a addition of impulse forces.
Particular solution of arbitrary force function $F(t)$ can therefore be expressed with the Green's function as
\be
 y_{\rm p}(t)  \; =\;  \int_{-\infty}^{t}  G(t-t_0) F(t_0)\, \d t_0\,,  \label{e13}
\ee
where the Green's function for the second-order LTI system is
\be
G(t-t_0) =   \f{1}{\o_{\rm d}}  e^{-\beta(t-t_0)} \sin \l(\o_{\rm d}(t-t_0)\r) \;\;\;{\rm for}\;\;\;  t \geq t_0 \,,  \label{e14}
\ee
otherwise zero (see textbooks, e.g. Marion and Thornton \cite{marion} or Kibble and Berkshire \cite{Kibble}). Here  $\o_{\rm d}  \equiv  \sqrt{\o_0^2  -  \beta^2} $.  The  Green's function found here can be applied to any second-order LTI systems in form of equation (\ref{e7}).

\begin{table*}[!h]
\begin{ruledtabular}
\begin{tabular}{cccc}
\textbf{Dynamical Parameters} & \textbf{Mechanical Vibration} & \textbf{Series RLC circuit} & \textbf{Parallel RLC circuit}  \\
\hline
 displacement  &  displacement,$\;y(t)$   & charge,$\;q(t)$ or current,$\;i(t)$   &  voltage,$\;v(t)$ \\
inertia  & $m$   & $L$   &  $C$ \\
resistance & $b$ & $R$   & $1/R $ \\
elasticity & $s$ & $1/C$  &  $1/L$ \\
$\o_0^2$ & $s/m $ & $1/(LC)$  &  $1/(LC)$ \\
$2 \beta$ & $ b/m $ & $R/L$  &  $1/(RC)$ \\
\end{tabular}
\caption{Dynamical parameters in second-order LTI systems for mechanical vibration, series RLC circuit and parallel RLC  circuit.} \label{tab1}
\end{ruledtabular}
\end{table*}

\section{Time-independent Schr\"{o}dinger Equation: Helmholtz equation}  \label{sec2}
 \subsection{Helmholtz equation}
In non-relativistic quantum mechanics, the
 Schr\"{o}dinger equation,
 \be
\f{- \hbar^2}{2 m} \nabla^2 \psi({\bf r})   +   V({\bf r}) \psi({\bf r}) \: = \:  E \psi({\bf r})  \,,
\ee
 describes time-independent system of a particle with mass, $m$. Wave function is only spatial-dependent,  i.e. $\psi = \psi(\bf r)$.
The equation can be re-arranged to Helmholtz equation  (see, e.g. Barton \cite{bar}, Griffiths \cite{gf}) or Schiff \cite{S})
\be
\l( \nabla^2    +   k^2 \r)    \psi({\bf r})  \:=\:    \f{2 m}{\hbar^2}  V({\bf r}) \psi({\bf r}) \:  \equiv  \: Q({\bf r})\,, \label{e16}
\ee
with $k^2 \equiv {2 m E}/{\hbar^2}$.  The LTI system can take similar form for $\beta = 0$ and with spatial dependency instead of temporal dependency.   If there is a response solution, $G({\bf r})$ to delta function  $\delta^3({\bf r})$ such that
\be
\l( \nabla^2    +   k^2 \r)   G({\bf r})  \:=\:   \delta^3({\bf r}) \,,   \label{e17}
\ee
then for an arbitrary inhomogeneous part $Q({\bf r})$ (the ``source"), one can find particular solution,
\be
\psi_{{\rm p}}({\bf r})\:=\:   \int_{-\infty}^{\bf r} G({\bf r} - {\bf r}_0)  Q({\bf r}_0) \d^3{\bf r}_0\,.   \label{e18}
\ee
The Green's function for the equation (\ref{e17}) has been known as,
\be
G({\bf r}) \: =  \:  -\f{e^{i k r}}{4 \pi r}\,.  \label{e20}
\ee
When inhomogenous part is not presented, i.e. $V({\bf r}) = 0$, the Green's function is $G_0({\bf r})$, hence
\be
\l( \nabla^2    +   k^2 \r) G_0({\bf r}) \,= \,0\,.   \label{e21}
\ee
Adding equations (\ref{e17}) and (\ref{e21}), hence
$
\l( \nabla^2    +   k^2 \r)\l[G({\bf r}) + G_0({\bf r}) \r] \,= \,  \delta^3({\bf r}). 
$
One can find complementary solution, $\psi_{\rm c}({\bf r})$ and general solution,
$ \psi({\bf r}) =\psi_{\rm c}({\bf r}) + \psi_{\rm p}({\bf r})\,,   $     
 of the system. Using the equations (\ref{e16}), (\ref{e18}) and (\ref{e20}), the general solution is therefore
\be
\psi({\bf r}) =\psi_{\rm c}({\bf r}) \, + \,  \f{- m}{2 \pi \hbar^2} \int_{-\infty}^{{\bf r}} \f{e^{ik | {\bf r}-{\bf r}_0 |}}{|{\bf r} - {\bf r}_0|}  V({\bf r}_0) \psi({\bf r}_0)\, \d^3 {\bf r}_0\,.
\label{e24}
\ee
The second term on the right-hand side is $\psi_{\rm p}({\bf r})$. This is integral form of the Schr\"{o}dinger equation. We consider  $\psi_{\rm c}({\bf r}) $ as a plane wave, $\psi_{0}({\bf r}) $ of an  incoming particle approaching a massive point of scattering at ${\bf r} =  {\bf r_0} $ with scattering potential $V({\bf r}_0)$. After scattering, $\psi_{\rm p}(t)$ is considered as {``response"} wave function at far distance from the scattering point.

\subsection{Born Approximation in quantum mechanics}
\begin{figure}
\centering
\includegraphics[width=5.8in]{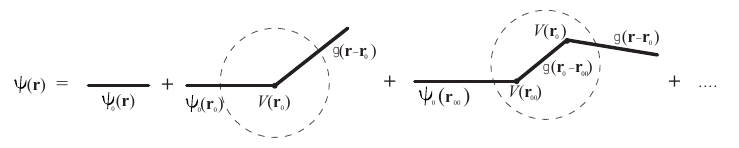}
\caption{Quantum mechanical-scattering Born series written as Feynman diagrams in two dimensions of space.} \label{F}
\end{figure}

The solution $\psi_{\rm p}(t)$ can be analyzed perturbatively. One well-known method is to use Born approximation in quantum mechanics (see e.g. Griffiths \cite{gf} and Schiff \cite{S}).  For simplicity, we express
$
g({\bf r})  \, \equiv   \,   -\l({m}/{2 \pi \hbar^2}\r)  \l({e^{ik r}}/{r}\r)
$, the equation (\ref{e24}) is hence
\be
\psi({\bf r}) \; = \;\psi_0({\bf r})  +   \int_{-\infty}^{{\bf r}} g({\bf r} - {\bf r}_0 )    V({\bf r}_0) \psi({\bf r}_0)\, \d^3 {\bf r}_0\,.
\label{e26}
\ee
Born approximation is to consider that at ${\bf r}_0$ the incoming plane wave is not much affected by the potential, i.e.
$
\psi({{\bf r}_0})\;   \approx \;  \psi_0({{\bf r}_0})  \,.
$
Hence the general solution is approximated as
\be
\psi({\bf r}) \; \approx \;\psi_0({\bf r})  +   \int_{-\infty}^{{\bf r}} g({\bf r} - {\bf r}_0 )    V({\bf r}_0) \psi_0({\bf r}_0)\, \d^3 {\bf r}_0\,.
\label{e28}
\ee
It is sensible to write down $\psi_0({\bf r}_0)$ as a scattered wave from ${\bf r}_{00}$ with the incoming wave $\psi_{00}({\bf r_0})$, hence
\be
\psi_0({\bf r}_0) \; = \;\psi_{00}({\bf r_0})  +   \int_{-\infty}^{{\bf r}_0} g({\bf r}_0 - {\bf r}_{00} )    V({\bf r}_{00}) \psi_0({\bf r}_{00})\, \d^3 {\bf r}_{00}\,.
\label{e29}
\ee
That is to say, the plane wave was scattered once at ${\bf r_{00}}$ by $V({\bf r}_{00})$ before arriving at  ${\bf r_{0}}$. Inserting equation (\ref{e29}) to   (\ref{e28}), we obtain
\bea
\psi({\bf r}) \; &\approx& \;\psi_{0}({\bf r}) \;    +\;   \int_{-\infty}^{{\bf r}} g({\bf r} - {\bf r}_{0} )    V({\bf r}_{0}) \psi_{00}({\bf r}_{0})\, \d^3 {\bf r}_{0}  \no \\
                    \; & & \; +  \int_{-\infty}^{{\bf r}} \int_{-\infty}^{{\bf r}_0} \Big[ g({\bf r} - {\bf r}_{0} ) V({\bf r}_0)\Big] \Big[  g({\bf r}_0 - {\bf r}_{00}) V({\bf r}_{00}) \Big] \psi_0({\bf r}_{00})    \;  \d^3 {\bf r}_{00}   \:    \d^3 {\bf r}_{0} \,.
\label{e30}
\eea
The wave function in the second term on the right-hand side, $\psi_{00}({\bf r}_{0})$ under Born approximation is
   $\psi_{00}({\bf r}_{0})  \approx \psi_{0}({\bf r}_{0})$, hence
\bea
\psi({\bf r}) \; &\approx& \;\psi_{0}({\bf r})  \hspace{10.34cm}(\rm 0^{th}\;order)   \no \\  \; & & \;  +\;   \int_{-\infty}^{{\bf r}} g({\bf r} - {\bf r}_{0} )    V({\bf r}_{0}) \psi_{0}({\bf r}_{0})\, \d^3 {\bf r}_{0}  \hspace{6cm} (\rm 1^{st}\;order)  \no \\
                    \; & & \; +  \int_{-\infty}^{{\bf r}} \int_{-\infty}^{{\bf r}_0} \Big[ g({\bf r} - {\bf r}_{0} ) V({\bf r}_0)\Big] \Big[  g({\bf r}_0 - {\bf r}_{00}) V({\bf r}_{00}) \Big] \psi_0({\bf r}_{00})    \;  \d^3 {\bf r}_{00}   \:    \d^3 {\bf r}_{0} \,.
            \hspace{0.73cm}(\rm 2^{nd}\;order)
\label{e31}
\eea
The first term is a plane wave, $\psi_{0}$ without scattering. In the second term, $\psi_{0}$  is scattered once at ${\bf r}_0$. The third term represents the incoming plane wave $\psi_{0}$ scattered twice, first at ${\bf r}_{00}$ and then at ${\bf r}_{0}$. We can extend this approximation beyond second order and the series becomes infinite series known as Born series.  This infinite series can be drawn in spirit of Feynman diagram as in Fig. \ref{F}.

\section{Born Approximation for an LTI system}   \label{sec3}
\begin{figure}
\centering
\includegraphics[width=6.5in]{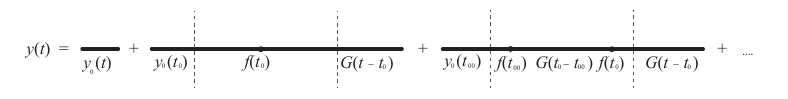}
\caption{Diagrams (in spirit of the Feynman graph) of  Born series for a second order LTI  system.}
 \label{F2}
\end{figure}

Here we apply the Born approximation directly to the second-order LTI system.  Let us start by considering the driving-force term $F(t)$-an inhomogeneous part, as a product of function $f(t)$ and the general solution $y(t)$,
\be
F(t) \: \equiv \:  f(t) y(t)\,.
\ee
The function $f(t)$ has the same role as the potential term,  $V({\bf r})$ in quantum mechanics. That is to say
  $f(t)$ represents external influence on the LTI system in similar manner to the effect of  $V({\bf r})$  in scattering problem.
  The second-order LTI system hence can be rewritten in form of Schr\"{o}dinger-like equation but with extra first-order derivative term $-2 \beta \d y/ \d t$ as
  \be
- \f{\d^2}{\d t^2} y - 2 \beta  \f{\d}{\d t}  y  +  f(t)y(t)   \:=\: \o_0^2 y    \label{e32_5}
 \ee
 The analogous quantities are
 \bea
 - \f{\d^2}{\d t^2} y(t)  \;& \Longleftrightarrow &\;  -\f{ \hbar^2}{2 m} \nabla^2 \psi({\bf r}) \no \\
  f(t)\, y(t) \; & \Longleftrightarrow &\;  V(\bf {r}) \,\psi(\bf {r}) \no \\
    \o_0^2\, y(t) \; & \Longleftrightarrow &\;  E \,\psi(\bf {r}) \no
 \eea
  In the limit of  $ \beta \rightarrow 0 $ (i.e., $b \rightarrow 0$ in mechanical LTI system,  $R  \rightarrow 0$ in series RLC and $R  \rightarrow \infty$ in paralell RLC circuits), the second-order LTI system and Schr\"{o}dinger equation are mathematically analogous to each other.
  At time $ t \geq  t_0$, the system is under influence of the driving force  $F(t)$. The general solution (\ref{e8}) is
  \be
y(t)  \;=\; y_{0}(t)  \:+\:      \int_{-\infty}^{t}  G(t-t_0) f(t_0)\, y(t_0)  \: \d t_0\,,  \label{e33}
\ee
 where the complementary solution $y_{\rm c}(t)$ is renamed to $y_{0}(t) $. Born approximation for the LTI case is
 $
 y(t_0) \: \approx \:  y_{0}(t_0)\,,
 $
which means at time $t=t_0$ the complementary solution is not much altered. Born series for LTI system is hence
\bea
y(t) \; &\approx& \; y_{0}(t)  \hspace{10.36cm}(\rm 0^{th}\;order)   \no \\
\; & & \;  +\;   \int_{-\infty}^{t} G(t - t_{0} )    f(t_{0}) y_{0}(t_{0})\, \d t_{0}  \hspace{6.293cm} (\rm 1^{st}\;order)  \no \\
                    \; & & \; +  \int_{-\infty}^{t} \int_{-\infty}^{t_0} \Big[ G(t - t_{0} ) f(t_0)\Big] \Big[  G(t_0 - t_{00}) f(t_{00}) \Big] y_0(t_{00})    \;  \d t_{00}   \:    \d t_{0}
            \hspace{1.272cm}(\rm 2^{nd}\;order)  \no \\   \; & & \; + \;\; {\rm higher\;order\; terms} \;\;+\; \; \ldots\,.
\label{e30}
\eea
The system is one dimension hence the Feynman graph is only straight line as seen in Fig. {\ref{F2}}. Let us consider application  in a problem of which the complementary and particular solutions are known, e.g. when the driving force is harmonic.

\subsection{Harmonic Driving Force}
Consider harmonic driving force, $F \:=\: F_0  e^{i \o t} $ in the LTI system.
 When the system is homogenous ($F = 0$), the complementary solution is
\be
y_{0}(t)\:=\: A e^{pt}\,,   \label{e28x}
\ee
where $A$ is a constant and
$
p =   - \beta \pm ( \beta^2   -  \o_0^2 )^{1/2}\,.
$
We will do Born approximation,
$
y(t_0) \: \approx \:  y_{0}(t_0)
$
in our analysis as in the Helmholtz case.
Depending on the $p$ value, there are three cases of complementary solution: light damping, heavy damping and critical damping.
In light damping case, $\beta < \o_0$ which corresponds to $p = -\beta + i \,\o_{\rm d}$ (choosing positive root, $\o_{\rm d}  = \sqrt{\o_0^2 - \beta^2}$). The complementary solution is
\be
 y_{0}(t) \:   =   \:   A e^{-\beta t \,+\, i \,\o_{\rm d} t} \,,
\ee
where $A = \exp{(\phi_0)}$ is a phase constant fixed by initial condition.  Using Born approximation, we write down $f(t_0)$ as
\be
f(t_0)   \:=\:  \f{F(t_0)}{y(t_0)}     \:\approx\:  \f{F(t_0)}{y_{0}(t_0)}  \: = \:   \f{F_0}{A} e^{\beta t_0}\, e^{i(\o - \o_{\rm d}) t_0}\,.
\ee
Hence the solution written as Born series,
\bea
y(t)  &\approx&  y_0(t)   +  \  \int^{t}_{-\infty} \l(  \f{F_0}{\o_{\rm d}} \r) e^{-\beta(t-t_0)}  \Bigg\{ \sin \l[\o_{\rm d}(t-t_0)\r]   e^{i \o t_0} \Bigg\} \: \d t_0   \no  \\
\:& &       +
 \int^{t}_{-\infty}    \int^{t_0}_{-\infty}  \f{1}{A}  \l(  \f{F_0}{\o_{\rm d}} \r)^2      e^{-\beta(t-t_0-t_{00})}   \Bigg\{ \sin \l[\o_{\rm d}(t-t_0)\r]\: \sin \l[\o_{\rm d}(t_{0}-t_{00}) \r]
 e^{i\l[ (\o -\o_{\rm d})t_0  +   \o t_{00}  \r]}   \Bigg\} \: \d t_{00}\,  \d t_{0}  + \ldots\,,
\eea
using Green's function in equation (\ref{e14}).
The second term (first order) manifests the transient state ``beats" between
two frequencies, $\o$ and $\o_{\rm d}$ if the driving force frequency, $\o$,  is close to $\o_{\rm d}$. Higher order terms represent more complex oscillation modes  
with less and less contribution.

\subsection{Condition for approximation}
 Beginning with the equation (\ref{e7}) with harmonic driving force,
the  complementary solution is the equation (\ref{e28x}). The  Born approximation is to be valid if at $t_0$ the solution is not much altered. This implies
$y(t_0) \approx y_{0}(t_0)$. In addition, the acceleration of the solution is very small at $t_0$, i.e. $\ddot{y}(t_0) \approx 0$ which is called {\it slow-roll} approximation. Applying the slow-roll and Born approximations to the equation (\ref{e7}), we obtain
\bea
 2 \beta Ae^{p t_0} p  +  \o_0^2 A e^{p t_0}     & \approx &  F_0 e^{i \o t_0}\,,   \no
\eea
giving a condition
\bea
\o_{\rm d}^2   +  \beta^2    &  \approx &     \f{F_0}{A}  e^{\beta t_0}\,,  \label{e50}
\eea
which depends much on the initial phase $ A = e^{\phi} $.  For light damping case, $\o_{\rm d}^2 = \o_0^2 - \beta^2 $ hence
\be
\o_0^2 \approx  \f{F_0}{A}  e^{\beta t_0}\,.
\ee
If the damping is critical, $\o_0 = \beta $, i.e. $\o_{\rm d} = 0$, hence
\be
\beta^2  \; \approx \;   \f{F_0}{A}  e^{\beta t_0}\,.
\ee
For heavy damping $ \beta > \o_0 $, therefore, $\o_{\rm d} = i \sqrt{\beta^2  - \o_0^2} $, from (\ref{e50}), we obtain
 $
\o_0^2 \approx  ({F_0}/{A})  e^{\beta t_0}
$
which is the same as light damping case. When the damping is very heavy, i.e. $\o_0 \ll \beta$,  hence  $\o_{\rm d}^2  \approx -\beta^2$ and
we get
$ F_0   \approx    0 $.
These conditions are equivalent to imposing of the slow-roll and Born approximations to the LTI system.

\section{Transformation of the LTI system to the Helmholtz equation}    \label{sec4}
A second order LTI system as a series RLC circuit has been found that, under a transformation,
\be
i(t) \: =\: I(t)\, e^{-t R/2L}  \: =\:   I(t)  \,e^{-\beta t}\,,
\ee
it can be transformed into the Helmholtz equation. As a result, Fourier and Laplace transforms are applied to derive transient solution with designed initial conditions (Sumichrast (2012) \cite{L}). Here we apply Born approximation to the LTI system in form of Helmholtz equation. The analysis is the same as the Born approximation in quantum mechanics. For the LTI system (\ref{e7}), with transformation $y \equiv \tilde{y} e^{-\beta t}$, the equation (\ref{e7}) becomes one-dimensional Helmholtz equation,
\be
\f{\d^2 \tilde{y}}{\d t^2} \,  +  \, \omega_{\rm d}^2 \,\tilde{y}     \;=\;   F(t)\,,   \label{ea44}
\ee
where $
\omega_{\rm d}^2\,  =  \,   \omega_0^2 - \beta^2  \, = \, {a_0}/{a_2} - \l[{a_1^2}/({4\, a_2^2}) \r]
$
as before and this is $\omega_{\rm d}^2\, = \, (1/LC) - R^2/(4L^2)$ for series RLC circuit.  The Green's function is the solution of
\be
\l( \f{\d^2 }{\d t^2}   +  \omega_{\rm d}^2  \r) G(t) \;= \;  \delta(t)\,.
\ee
Defining $F(t)  =   \tilde{f}(t) \tilde{y}(t)$, the Schr\"{o}dinger-like form of this equation is now without
 damping term in comparison  to (\ref{e32_5}),
  \be
- \f{\d^2}{\d t^2} \tilde{y}   +   \tilde{f}(t)\tilde{y}(t)   \:=\: \o_{\rm d}^2 \tilde{y}     \,.
 \ee
The Green's function for this case is one-dimensional,
\be  G(t-t_0) \: = \:   - \f{ i e^{i \o_{\rm d}(t-t_0)}}{2 \o_{\rm d}}\,,  \ee
and the solution is
\be
\tilde{y}(t)  \:=\:   \tilde{y}_{\rm c} (t)\,  +\,  \int_{-\infty}^{t} G(t-t_0)  \tilde{f}(t_0) \tilde{y}(t_0)\, \d t_0 \,.
\ee
Using Born approximation $\tilde{y}(t_0)  \approx  \tilde{y_0}(t_0)   $, the Born series for the LTI system (in form of one-dimensional Helmholtz equation) is
\bea
\tilde{y}(t) \; &\approx & \; \tilde{y}_{0}(t)  \hspace{10.36cm}(\rm 0^{th}\;order)   \no \\
\; & & \;  +\;   \int_{-\infty}^{t} G(t - t_{0} )    \tilde{f}(t_{0}) \tilde{y}_{0}(t_{0})\, \d t_{0}  \hspace{6.293cm} (\rm 1^{st}\;order)  \no \\
                    \; & & \; +  \int_{-\infty}^{t} \int_{-\infty}^{t_0} \Big[ G(t - t_{0} ) \tilde{f}(t_0)\Big] \Big[  G(t_0 - t_{00}) \tilde{f}(t_{00}) \Big] \tilde{y}_0(t_{00})    \;  \d t_{00}   \:    \d t_{0} \,.
            \hspace{1.272cm}(\rm 2^{nd}\;order)  \no \\   \; & & \; + \;\; {\rm higher\;order\; terms} \;\;+\; \; \ldots\,   \label{eLTIh}
\eea
Now let us consider a case of harmonic driving force $F(t)  =  F_0 e^{i \o t}$, for the equation (\ref{ea44}), the complementary solution is
\be
\tilde{y}_0(t)  \, = \,  B e^{i \o_{\rm d} t}\,,
\ee
where $B$ is a constant. Hence from $F(t)  =   \tilde{f}(t) \tilde{y}(t)$ we have $F(t_0) \approx  \tilde{f}(t_0) \tilde{y}_0(t_0)$ and $\tilde{f}(t_0)  =  (F_0/B) e^{i(\o - \o_{\rm d}) t_0}$.
Finally the Born series (\ref{eLTIh}) for the case of harmonic driving force is written as
\bea
\tilde{y}(t) \; &\approx & \; \tilde{y}_{0}(t)  \hspace{10.36cm}(\rm 0^{th}\;order)   \no \\
\; & & \;  +\;   \int_{-\infty}^{t}
\l(\f{- i F_0}{2 \o_{\rm d}} \r)   \l[   \,e^{i \o_{\rm d}(t-t_0) + i \o t_0}   \r]
    \, \d t_{0}  \hspace{5.345cm} (\rm 1^{st}\;order)  \no \\
                    \; & & \; +  \int_{-\infty}^{t} \int_{-\infty}^{t_0}     \l( \f{-i F_0}{2 \o_{\rm d}}\r)^2  \l( \f{ 1  }{B}  \r)
                      \l[  e^{i \o_{\rm d} (t - t_0 - t_{00}) }    \, e^{i \o (t_0 + t_{00})}    \r]        \;  \d t_{00}   \:    \d t_{0} \,.
            \hspace{1.272cm}(\rm 2^{nd}\;order)  \no \\   \; & & \; + \;\; {\rm higher\;order\; terms} \;\;+\; \; \ldots\,.
\eea

\section{conclusion}
  In this work, we show a mathematical connection of the LTI and the Helmholtz equation. We show that it is valid to apply the Born approximation to the LTI system. Although it might not be very useful in practice for experiment on RLC system since it is not difficult to set up the initial value of input signal, the approximation could be useful when the RLC circuit is not the case, for example, the case of mechanical oscillation.
The Born approximation in quantum mechanics can be used to find particular solution of the LTI oscillations. With the Born approximation, we can approximate that, when force starts to exert on the system, the solution is not much altered from the complementary solution. This gives us some alternative analytic way of tackling the problem. Moreover, we express the solution in Born series as graphical term in spirit of Feynman diagrams. For a case study of harmonic forced oscillation, conditions corresponding to the Born approximation with slow-roll assumption are derived.   Born series of harmonic forced oscillation case are also expressed either by directly applying the approximation to the LTI system or transforming the LTI system to Helmholtz form before applying the approximation.
We comment that in quantum mechanics if one can 
express the source term $Q({\bf r})$ of Helmholtz equation explicitly with its boundary value, the particular solution can be found using Green function method and there is no need to do Born approximation.

\section*{Acknowledgements}
This project is funded by the National Research Council of Thailand (Grant: R2554B072). The author thanks Pramote Wardkien (Department of Telecommunication, KMITL)  and Seckson Sukhasena for discussions.

\end{document}